\documentclass[twoside,twocolumn,english,british,prb, showpacs]{revtex4}
\usepackage[latin9]{inputenc}
\usepackage{float}
\usepackage{amstext}
\usepackage{amssymb}
\usepackage{graphicx}
\usepackage{esint}

\makeatletter
\@ifundefined{textcolor}{}
{%
 \definecolor{BLACK}{gray}{0}
 \definecolor{WHITE}{gray}{1}
 \definecolor{RED}{rgb}{1,0,0}
 \definecolor{GREEN}{rgb}{0,1,0}
 \definecolor{BLUE}{rgb}{0,0,1}
 \definecolor{CYAN}{cmyk}{1,0,0,0}
 \definecolor{MAGENTA}{cmyk}{0,1,0,0}
 \definecolor{YELLOW}{cmyk}{0,0,1,0}
}

\makeatother

\usepackage{babel}
\begin{document}

\title{Luttinger parameters of interacting fermions in 1D at high energies}

\author{O. Tsyplyatyev}

\affiliation{School of Physics and Astronomy, The University of Birmingham, Birmingham,
B15 2TT, UK}

\author{A. J. Schofield }

\affiliation{School of Physics and Astronomy, The University of Birmingham, Birmingham,
B15 2TT, UK}

\date{\today}
\begin{abstract}
Interactions between electrons in one-dimension are fully described
at low energies by only a few parameters of the Tomonaga-Luttinger
model which is based on linearisation of the spectrum. We consider
a model of spinless fermions with a short range interaction via the
Bethe-Ansatz technique and show that a Luttinger parameter emerges
in an observable beyond the low energy limit. A distinct feature of
the spectral function, the edge that marks the lowest possible excitation
energy for a given momentum, is parabolic for arbitrary momenta and
the prefactor is a function of the Luttinger parameter, $K$. 
\end{abstract}


\maketitle

\section{Introduction}
The effects of interactions between fermions in one-dimension are
mainly understood at low energies within the scope of the Tomonaga-Luttinger
model.\cite{giamachi} This framework is based on the linear approximation
to the single-particle spectrum around the Fermi energy and provides,
via the bosonisation technique,\cite{giamachi} a generic way to
calculate various correlation functions. Understanding of interacting
fermions beyond the low-energy limit still presents a challenge. Studies
are currently focused on dynamical response functions, \cite{Pustilnik06,Pereira06,Imambekov08,Essler10,JS05,Khodas07}
\textit{e.g.} the spectral function which can be measured by momentum
resolved tunnelling of electrons in semiconductors, \cite{Auslaender,Ritchie}
by angle resolved photo-emission in correlated materials, \cite{Maekawa}
and by photoemission spectroscopy in cold atoms.\cite{Stewart}  Recently
significant theoretical progress was achieved in this direction by
making a connection between Luttinger liquids and the Fermi edge singularity
problem.\cite{GlazmanReview} As a result power-law singularities
were found at the edge of the spectral function at zero temperature
and their powers were related to the corresponding curvature. \cite{ImambekovGlazman}
The edge marks the smallest energy at a fixed momentum with which
a particle can tunnel into the system. At low energies the edge disperses
linearly with a slope which is the sound velocity of collective modes
$v$ defined by parameters from the Tomonaga-Luttinger model; \cite{LutherPeschel}
a small quadratic correction to the linear slope at low momenta was
found in Ref. \onlinecite{Pereira06}. In this paper we calculate the position
of the edge for the spinless fermions with a short range interaction
at arbitrary energies and show that a Luttinger parameter is still
relevant at large energies. 

Our strategy is to consider the exact diagonalisation of the model
on a lattice via the Bethe-Ansatz approach. Then we analyse the spectral
function in the continuum regime -- a combination of the thermodynamic
limit and a small occupancy of the lattice\cite{ContinuumRegime}
-- which corresponds to the continuum model with a contact interaction.
In this regime we find that the position of the edge is parabolic
for arbitrary momenta and the prefactor is a function of the dimensionless
Luttinger parameter $K$ {[}see Eq. (\ref{eq:eedge}){]} which is
defined in the low energy domain of the Tomonaga-Luttinger model.
Our result could be directly observed in experiments on spin-polarised
particles such as electrons in ferromagnetic semiconductors \cite{ohno}
using the setups of Refs. \onlinecite{Auslaender,Ritchie} or polarised
cold atoms using the setup of Ref. \onlinecite{Stewart}. In closely related models of spin chains\cite{giamachi} the position of the edge depends on the Luttinger $K$ in an analogous way but, for example for a weakly polarised chain, the parabolic function of momentum becomes a cosine. With the parabolic shape found in
this paper, the phenomenological non-linear Luttinger liquid theory\cite{ImambekovGlazman} gives a divergent power of the edge singularity.

In the continuum regime the Luttinger parameter, $K$, is bounded
and the smallest $K$ for large interaction strengths is almost degenerate
with its non-interacting value $K=1$. We use the Bethe-Ansatz approach
for a finite-range interaction potential beyond nearest neighbour
in the limit $V=\infty$ and show that the regime of strong interaction effects (corresponding
to the minimum value of $K=0$ in the Tomonaga-Luttinger  model) can
only be accessed by a microscopic model with the interaction range
at least of the order of the average distance between particles.

The paper is organised as follows.  Section II contains definition of the model of spinless femions on a lattice and the spectral function. In Section III we analyse the edge of the spectral function for next-neighbour interaction in the low (Subsection A) and high (Subsection B) regimes.  In Section IV we consider a finite range interaction in the limit of infinite interaction strength. In appendix we give numerical data that clarify calculation in Sections III and IV.

\section{Model}
Spinless fermions on a one-dimensional lattice with $L$ sites interacting
via a two body-potential, $V_{i}$, as
\begin{equation}
H=-t\sum_{j=1}^{L}\left(c_{j}^{\dagger}c_{j+1}+c_{j}^{\dagger}c_{j-1}\right)+\sum_{j=1,i=1}^{L,\infty}V_{i}c_{j}^{\dagger}c_{j}c_{j+i}^{\dagger}c_{j+i}\label{eq:H}
\end{equation}
where $t$ is a hopping amplitude and operators $c_{j}$ obey Fermi
commutation relations $\left\{ c_{i},c_{j}^{\dagger}\right\} =\delta_{ij}$.\cite{hbara} Below we consider periodic boundary conditions $c_{L+1}=c_{1}$
to maintain the translation symmetry of the finite length chain and
consider only repulsive interactions, $V_{i}>0$.

The spectral function describes the tunnelling probability for a particle
with momentum $k$ and energy $\varepsilon$, $A\left(k,\varepsilon\right)=-\textrm{Im}G\left(k,\varepsilon\right)\textrm{sgn\ensuremath{\left(\varepsilon-\mu\right)}/\ensuremath{\pi}}$
where $\mu$ is the chemical potential and $G\left(k,\varepsilon\right)=-i\sum_{j}\int dte^{i\left(kj-\varepsilon t\right)}\left\langle T\left(e^{-iHt}c_{j}e^{iHt}c_{1}\right)\right\rangle /L$
is a Fourier transform of the single particle Green function at zero
temperature. To be specific we discuss only a particular region, $k_{F}<k<3k_{F}$
and $\varepsilon>\mu$. The spectral function in this domain reads
\cite{AGD}

\begin{equation}
A\left(k,\varepsilon\right)=L\sum_{f}\left|\left\langle f\left|c_{1}^{\dagger}\right|0\right\rangle \right|^{2}\delta\left(k-P_{f}\right)\delta\left(\varepsilon+E_{0}-E_{f}\right),\label{eq:Ake}
\end{equation}
where $E_{0}$ is the energy of the ground state $\left|0\right\rangle $,
$P_{f}$ and $E_{f}$ are the momenta and the eigenenergies of the
eigenstates $\left|f\right\rangle $; all eigenstates are assumed
normalised. 

\section{Next-neighbour interaction}
The model of Eq. (\ref{eq:H}) can be diagonalised using the Bethe
Ansatz when the interaction potential is restricted to the nearest
neighbour only, $V_{i}=V\delta_{i,1}$.\cite{KBI} In the coordinate
basis, $\left|\psi\right\rangle =\sum_{j_{1}<\dots<j_{n}}a_{j_{1}\dots j_{n}}c_{j_{1}}^{\dagger}\dots c_{j_{n}}^{\dagger}\left|\textrm{vac}\right\rangle $
where $\left|\textrm{vac}\right\rangle $ is the fermionic vacuum,
a superposition of plain waves $a_{j_{1\dots}j_{n}}=\sum_{P}e^{i\sum_{l=1}^{n}k_{P_{l}}j_{l}+i\sum_{l<m=1}^{n}\varphi_{P_{l},P_{m}}}$
is an $n$ particle eigenstate, $H\left|\psi\right\rangle =E\left|\psi\right\rangle $,
with the eigenenergy 
\begin{equation}
E=-2t\sum_{j=1}^{n}\cos\left(k_{j}\right)+2tn\;.\label{eq:E}
\end{equation}
Here a constant $2tn$ was added for convenience, the phase shifts

\begin{equation}
e^{i2\varphi_{jm}}=-\frac{e^{i\left(k_{j}+k_{m}\right)}+1+\frac{V}{t}e^{ik_{j}}}{e^{i\left(k_{j}+k_{m}\right)}+1+\frac{V}{t}e^{ik_{m}}}\;.
\end{equation}
are fixed by the two-body scattering problem and $\sum_{P}$ is a
sum over all permutations of $n$ integer numbers. The periodic boundary
condition quantises all single particle momenta simultaneously, 

\selectlanguage{english}%
\begin{equation}
Lk_{j}-2\sum_{m}\varphi_{jm}=2\pi\lambda_{j},\label{eq:BAE}
\end{equation}
where $\lambda_{j}$ are integer numbers. The sum $P=\sum_{j}k_{j}$
is a conserved quantity---the total momentum of an $n$ particle state.

The solutions of the non-linear system of equations Eq. (\ref{eq:BAE})
can be classified in the limit of non-interacting particles. Under
substitution of the scattering phase $2\varphi_{jm}=\pi$ for $V=0$
Eq. (\ref{eq:BAE}) decouples into a set of independent quantisation
conditions for plain waves, 
\begin{equation}
k_{j}=\frac{2\pi\lambda_{j}}{L}.\label{eq:k0}
\end{equation}
The corresponding eigenstates are Slater determinants which vanish
when the momenta of any two particles are equal. Thus all eigenstates
are mapped onto all possible sets of $n$ non-equal integer numbers
$\lambda_{j}$ with $-L/2<\lambda_{j}\leq L/2$. In the absence of
bound state formation, these solutions are adiabatically continued
under a smooth deformation from $V=0$ to any finite value of $V$.
This permits us to use the free particle classification to label many-particle
states for an arbitrary interaction strength.

The limit of infinitely strong repulsion corresponds to free fermions
of a finite size. The scattering phase $\varphi_{jm}=k_{j}-k_{m}+\pi$
for $V=\infty$ makes Eq. (\ref{eq:BAE}) a linear system of coupled
equations. In the continuum regime they decouples into a set of single
particle quantisation conditions, 

\selectlanguage{british}%
\begin{equation}
k_{j}=\frac{2\pi\lambda_{j}}{L-n}.\label{eq:k1inf}
\end{equation}
Here the length of the system is reduced
by the exclusion volume taken by the finite size of the particles, see also Eq. (\ref{eq:kjr}) for a finite range interaction below.

The adiabatic method we are using breaks down \foreignlanguage{english}{when
a bound state is formed at a finite interaction strength while sweeping
from $V=0$ to $V=\infty$. Such states occur only when some of the
quasimomenta of the solutions at $V=0$ are }$\left|k_{j}\right|>\pi/2$, see appendix and Ref. \onlinecite{pereira09}. The bound states can be observed,
for instance, in dynamics of a spin chain following a quench.\cite{Essler12}
In the continuum regime there is a wide range of model parameters
where Eq. (\ref{eq:k1inf}) is applicable: for momenta and energies
in the spectral function smaller than $\pi/2$ and smaller than half
bandwidth respectively.

\begin{figure}
\centering\includegraphics[width=1\columnwidth]{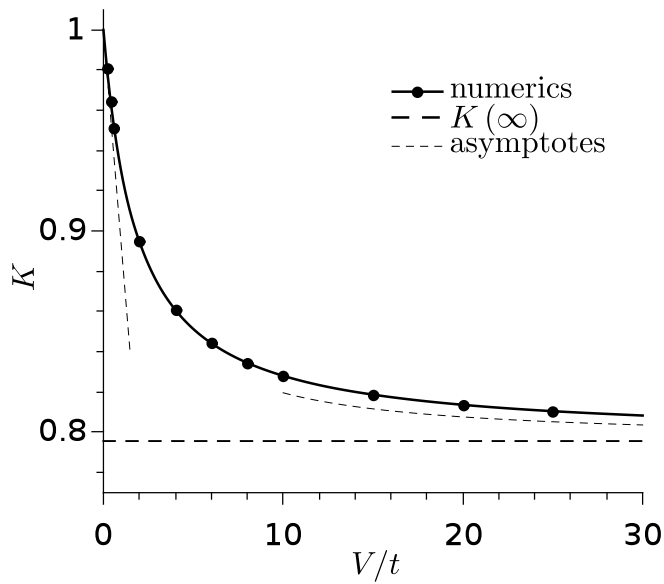}

\caption{\label{fig:K}The numerical evaluation of the Luttinger parameter
$K$ as a function of next-neighbour interaction strength $V$ using
Eqs. (\ref{eq:E}, \ref{eq:BAE}, \ref{eq:v}) - full ellipses, $L=100$
and $n=11$. We compare it with the bound on $K$ at infinite interaction
strength from Eq. (\ref{eq:K1inf}) - thick dashed line. Small and
larger $V$ asymptotes are $K=1-2n/L\times V/t$ and $K=K\left(\infty\right)+2n/\left(L-n\right)\times t/V$
- thin dashed lines.}
\end{figure}
The ground state is a band filled from the bottom up to the momentum
$k_{F}=\pi\left(n-1\right)/L$ using the classification of Eq. (\ref{eq:k0}).
Here $n$ is assumed odd for simplicity. Eigenstates involved in the
form factors of the spectral function have a fixed number of particles
$n+1$. All other eigenstates do not contribute to Eq. (\ref{eq:Ake})
as the number of particles is a conserved quantity.

In this paper, we are concerned with the location of the support of
the spectral function (the lowest value of energy for which the spectral
function is not zero) as opposed to its value so we ignore the matrix
elements in Eq. (\ref{eq:Ake}) assuming them to be non-zero for all
$f$ which satisfy the number constraint. Two delta-functions in $k$
and in $\varepsilon$ map directly the total momenta and the eigenenergies
of all many-body states $\left|f\right\rangle $ with $n+1$ particles
into the shape of the spectral function. For a fixed value of $k$,
the edge of the support is the smallest eigenenergy of all states
$\left|f\right\rangle $ with $P_{f}=k.$ Using the classification
in Eq. (\ref{eq:k0}) these states can be parameterised by a single
variable, $\Delta P$, see the sketch in Fig. \ref{fig:edge}(b) and
in appendix.

\subsection{Low energies}
At low energies the model of spinless fermions Eq. (\ref{eq:H}) is
well approximated by the Tomonaga-Luttinger model with only two free
parameters.\cite{giamachi} The first parameter is the slope of the
linearised spectrum of excitations at $k_{F}$. For the states from
Fig. \ref{fig:edge}(b) it is 
\begin{equation}
v=\frac{L\left(E_{2}-E_{1}\right)}{2\pi},\label{eq:v}
\end{equation}
where $E_{1}$ and $E_{2}$ are energies of the states with $\Delta P=0$
and $\Delta P=2\pi/L$ respectively. The second Luttinger parameter
can be extracted as $K=v_{F}/v$, where $v_{F}=2t\pi\left(n-1\right)/L$
is the Fermi velocity of the non-interacting system. The numerical
evaluation of $K$ as a function of the interaction strength $V$
is presented on Fig.~\ref{fig:K}. For small $V$ the function is
linear, $K=1-2n/L\times V/t+O\left(V^{2}/t^{2}\right)$. For large
$V$ it approaches a lower bound such that $K=K\left(\infty\right)+2n/\left(L-n\right)\times t/V+O\left(t^{2}/V^{2}\right)$
where 
\begin{equation}
K\left(\infty\right)=\left(1-\frac{n}{L}\right)^{2},\label{eq:K1inf}
\end{equation}
was computed using the values of quasimomenta for $V=\infty$ in Eq.
(\ref{eq:k1inf}).

The Luttinger parameter, $K$, measures the effects of interactions
where for non-interacting particles $K=1$. At $V=\infty$ the interaction
potential is a hard wall interaction with a finite interaction range
which still leaves some room for non-zero kinetic energy thus limiting
the maximum value of $K$.

\subsection{High energies}
\begin{figure}
\centering\includegraphics[width=1\columnwidth]{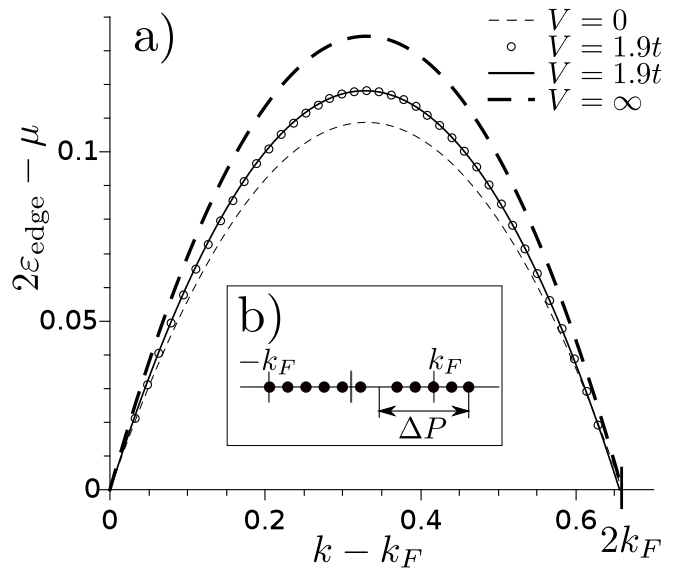}

\caption{\label{fig:edge}a) We show the main result of this paper - that the
edge of the support of the spectral function satisfies Eq. (\ref{eq:eedge}).
Numerical results at intermediate coupling ($V=1.9t$) are shown as
open circles and compared to the analytical result, Eq. (\ref{eq:eedge}),
shown as full line for $L=400$ and $n=39$. The asymptotes in the
weak coupling limit from Eq. (\ref{eq:k0}) and the strong coupling
limit from Eq. (\ref{eq:k1inf}) are shown as thin and thick dashed
lines respectively. b) Sketch of the sets of quasimomenta, using classification
Eq. (\ref{eq:k0}), that correspond to the edge states, parameter
$\Delta P$ corresponds to different momenta $k$.}
\end{figure}

The main aim of this paper is a calculation beyond low energies. In
the non-linear region the position of the edge of the spectral function
is given by the momentum dependence of the states of Fig. \ref{fig:edge}(b),
$\varepsilon_{\textrm{edge}}\left(k\right)=E_{k}-E_{0}$ where $E_{k}$
correspond to the states with $\Delta P=k_{F}+2\pi/N-k$. For all
values of $V$ we find it to be a parabolic function of momentum,

\begin{equation}
\varepsilon_{\textrm{edge}}\left(k\right)=\frac{mv_{F}^{2}}{K}-\frac{\left(k-2mv_{F}\right)^{2}}{2mK}\label{eq:eedge}
\end{equation}
where $m=\left(2t\right)^{-1}$ is the bare single electron mass and
the Luttinger parameter $K$ is determined by the slope at $k=k_{F}$.
In the limiting cases $V=0$ and $V=\infty$, it is calculated explicitly
using the expressions for quasimomenta in Eqs. (\ref{eq:k0}) and
(\ref{eq:k1inf}). The crossover for intermediate values of $V$ is
calculated using the numerical solution of the Bethe equations, Eq.
(\ref{eq:BAE}), and is perfectly fitted by the same parabolic formula,
see Fig. \ref{fig:edge}(a). 
 At $k=k_{F}$ Eq. (\ref{eq:eedge}) gives the chemical potential
$\mu=mv_{F}^{2}/\left(2K\right)$ since the ground state for $n+1$
particles is constructed by adding an extra particle to the ground
state $\left|0\right\rangle $ at the lowest possible momentum above
$k_{F}$ which is the state in Fig. \ref{fig:edge}(b) with $\Delta P=0$.

The many-body states that mark the edge of the spectral function outside
of the region $k_{F}<k<3k_{F}$ are parameterised by a single variable
similarly to Fig. \ref{fig:edge}(b) (see appendix
for details). In the upper half of the energy-momentum plane, $\varepsilon>\mu$,
the result in Eq. (\ref{eq:eedge}) is repeated along the momentum
axis with the period $2k_{F}$. So $\varepsilon_{\textrm{edge}}\left(k\right)$
becomes $mv_{F}^{2}/K-\left(k-2jmv_{F}\right)^{2}/\left(2mK\right)+\Delta\mu_{j}$
in regions $\left(2j-1\right)$$k_{F}<k<\left(2j+1\right)k_{F}$ with
an additional shift $\Delta\mu_{j}$ for $\left|j\right|>1$. The
latter is given by the recurrence relation $\Delta\mu_{j+1}=\Delta\mu_{j}+2\left|j\right|v_{F}/K$
with the initial value $\Delta\mu_{1}=0$. In the continuum regime
of our interest, $j\ll n$, $\Delta\mu_{j}$ is only a small finite
size correction to $\mu$. In the ``hole region'', $\varepsilon<\mu$,
the position of the edge is obtained by reflection of $\varepsilon_{\textrm{edge}}\left(k\right)$
with respect to the line $\varepsilon=\mu$, see appendix. 

A link between Luttinger liquids and the Fermi-edge singularity problem
was very recently established as a tool to analyse interactions beyond
the linear approximation in one-dimension.\cite{GlazmanReview} This
has led to the development of a phenomenological theory of non-linear
Luttinger liquids where power-law singularities, $A\left(\varepsilon,k\right)\sim\theta\left(\varepsilon-\varepsilon_{\textrm{edge}}\left(k\right)\right)\left|\varepsilon-\varepsilon_{\textrm{edge}}\left(k\right)\right|^{-\alpha}$,
were found above the edge of the support. Their exponents were related
to the curvature of \foreignlanguage{english}{$\varepsilon_{\textrm{edge}}\left(k\right)$
}for arbitrary momenta.\cite{GlazmanReview}
Substitution of Eq. (\ref{eq:eedge}) in the formula of Imambekov
and Glazman from Ref. \onlinecite{ImambekovGlazman} yields
\begin{equation}
\alpha=1-\frac{K}{2}\left(1-\frac{1}{K}\right)^{2}
\end{equation}
The Luttinger $K$ of the model Eq. (\ref{eq:H}), see Fig. \ref{fig:K},
gives a divergent exponent smaller than one and larger then a limiting
value calculated for $K\left(\infty\right)$ from Eq. (\ref{eq:K1inf}).
Thus the form factors in Eq. (\ref{eq:Ake}) are non-zero around the
edge thereby justifying our assumption about matrix elements in the
spectral function.

\section{Finite range interaction}
A further consequence of the non-linearity of the free particle dispersion
is the bound on the Luttinger parameter $K$ in Eq. (\ref{eq:K1inf}).
It has to be treated with care analogously to the point-splitting
technique for field theoretical models\cite{vonDelftSchoeller} in
which a small interaction range must be introduced to couple a pair
of fermions which cannot occupy the same point in space, then the
limit of zero range is taken. For the model on a lattice with next-neighbour
coupling, the interaction range vanishes in the continuum regime ($n\ll L$)
compared to the average distance between particles, therefore $K\left(\infty\right)\rightarrow1$
(\textit{i.e.} degenerate with its value for the non-interacting system
$K\left(0\right)=1$). However, the interaction range between fermions
in physical systems is usually finite, \textit{e.g.} the screening
length for electrons in a metal or a semiconductor, making $K$ not
equal to one. We, therefore, now consider a model with finite range.

We consider the limiting case of $V=\infty$ when the interaction
range (screening length) spans a large number of lattice sites, $r$.
The Hamiltonian Eq. (\ref{eq:H}) with the potential $V_{i}=V\theta\left[i-1\right]\theta\left[r-i\right]$,
where $\theta\left[i\right]$ ($\theta\left[i\right]=1$ for $i\geq0$
and $\theta\left[i\right]=0$ for $i<0$) is a Heaviside step function
and $V\rightarrow\infty$, can be diagonalised in the coordinate basis,
$\left|\psi\right\rangle =\sum_{j_{1}<j_{2}-r\dots<j_{n}-r}a_{j_{1}\dots j_{n}}c_{j_{1}}^{\dagger}\dots c_{j_{n}}^{\dagger}\left|\textrm{vac}\right\rangle $,
by a superposition of plain waves, $a_{j_{1\dots}j_{n}}=\sum_{P}e^{i\sum_{l=1}^{n}k_{P_{l}}j_{l}+i\sum_{l<m=1}^{n}\varphi_{P_{l},P_{m}}}$,
with $2\varphi_{jm}=\left(k_{j}-k_{m}\right)r+\pi$. Application of
the periodic boundary condition yields, similarly to Eq. (\ref{eq:BAE}),
\begin{equation}
k_{j}\left(L-r\left(n-1\right)\right)+r\sum_{m=1\neq j}^{n}k_{m}=2\pi\lambda_{j},
\label{eq:kjr}\end{equation}
which in the continuum regime gives a set of independent quantisation
conditions $k_{j}=2\pi\lambda_{j}/\left(L-rn\right)$. Finally, repeating
the same calculation used to obtain Eq. (\ref{eq:K1inf}) we find

\begin{equation}
K\left(\infty\right)=\left(1-\frac{rn}{L}\right)^{2},
\end{equation}
where the term $rn/L$ can be interpreted as a product of a screening
length and a particle density.

A microscopic model of spinless fermions needs to have an interaction
range of the order of the average distance between particles to reach
the $K=0$ value that corresponds to strong interaction effects in
the Tomonaga-Luttinger model. Specifically, for $r=L/\left(2n\right)$,
which allows some motion even when $V=\infty$, the bound is $K\left(\infty\right)=1/4$.
When $r$ is increased further, $K\left(\infty\right)$ approaches
zero.

\section{Conclusions}
In conclusion, we have considered the exact diagonalisation of a model
of spinless fermions on a lattice with next-neighbour interactions
via the Bethe-Ansatz approach. Analysing the spectral function in
the continuum regime we have found that the edge of its support has
a parabolic shape for arbitrary momenta and the prefactor is a function
of the dimensionless Luttinger parameter $K$ which is defined in
the low energy domain. Additionally we have extended our model with
a finite range of the interactions in order to access the strongly
interacting regime (near $K=0$) and have also found the parabolic
shape for the support (for $V=\infty$) which is still characterised
by $K$. This suggests that Luttinger parameters control physical
properties at higher energies where the non-linearity cannot be ignored.

\section{Acknowledgement}
We would like to thank I. V. Lerner for illuminating discussions.
This work was supported by EPSRC grant EP/J016888/1.

\appendix*

\section{Numerical data}
Here we present results of numerical calculations. Fig. 1 shows some of the solutions to the Bethe-Ansatz
equations Eq. (5) of the main text for different values of $V$. The
states are parameterised using Eq. (6) of the main text. The states
on Fig.1(a)-(c) have all $\left|k_{j}\right|<\pi/2$. The state on
Fig. 1(d) contains a pair of $\left|k_{j}\right|>\pi/2$ that leads
to formation of a bound state at a finite $V$. Fig. 2 shows the extension
of the edge beyond the region $k_{F}<k<3k_{F}$ and $\varepsilon>\mu$.
The eigenstates on the edge are marked by large dots and corresponding
sets of quasimomenta are sketched in each region as insets.

\onecolumngrid
 
\begin{figure}[H]
\centering\includegraphics[width=0.5\columnwidth]{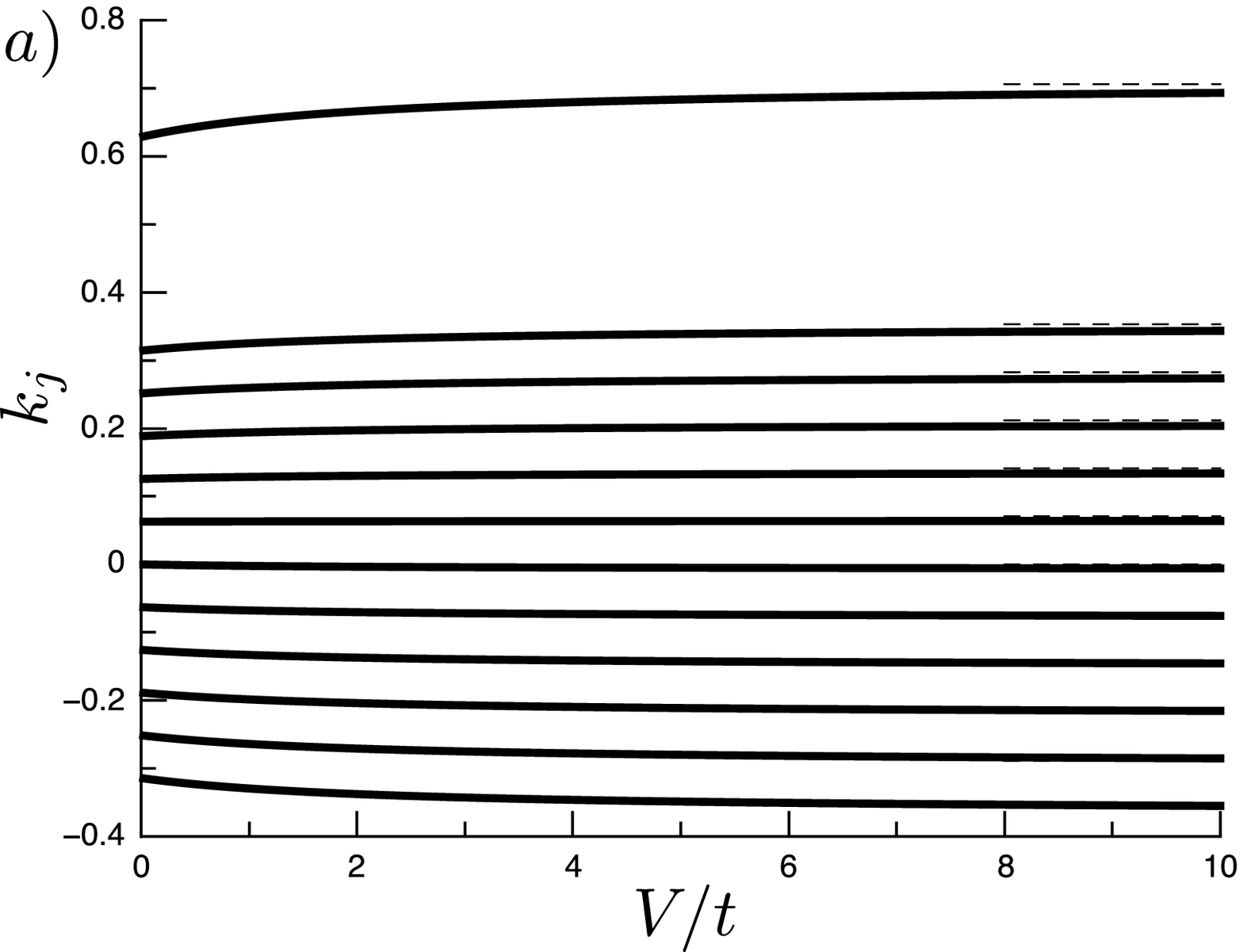}\includegraphics[width=0.5\columnwidth]{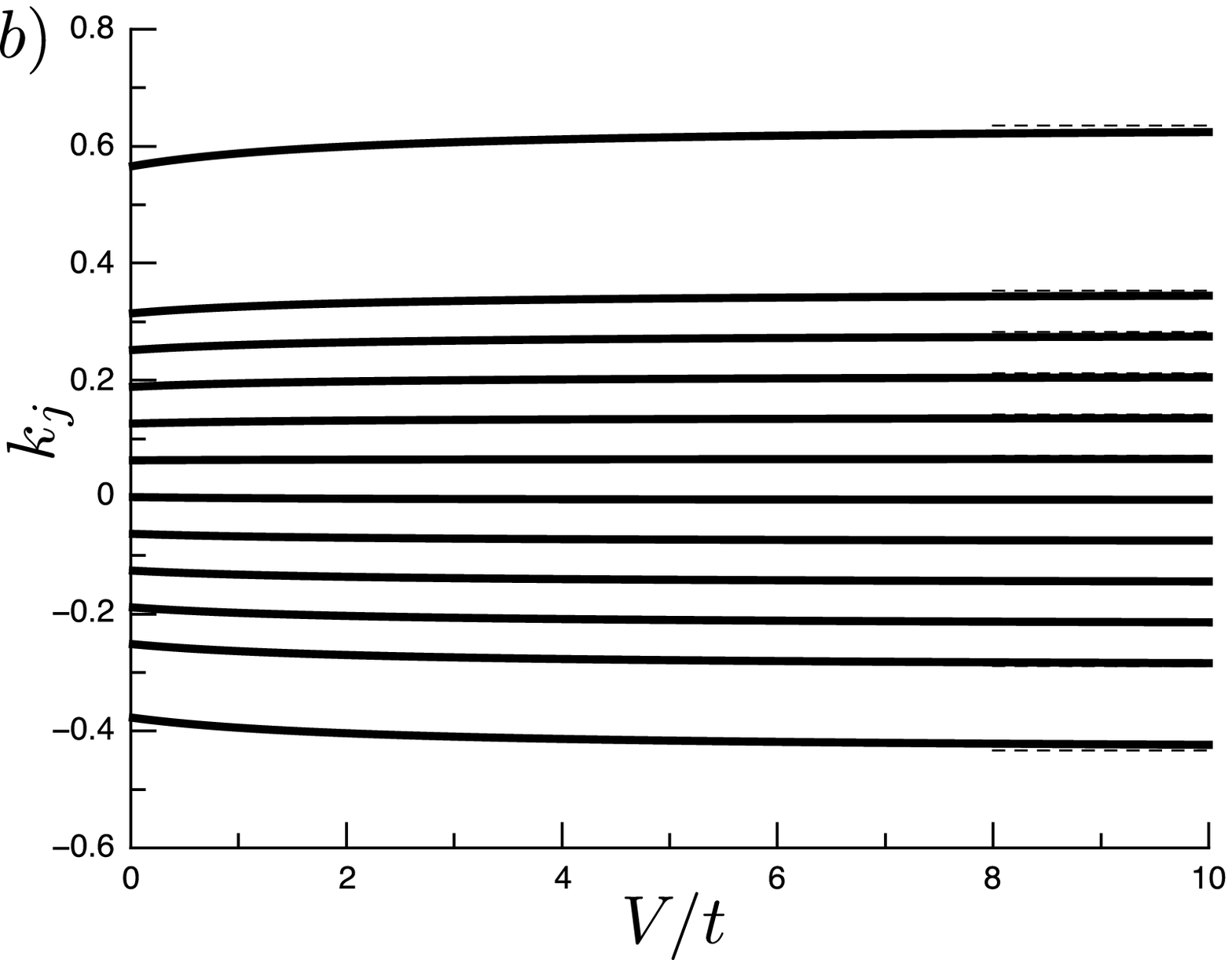}

\includegraphics[width=0.5\columnwidth]{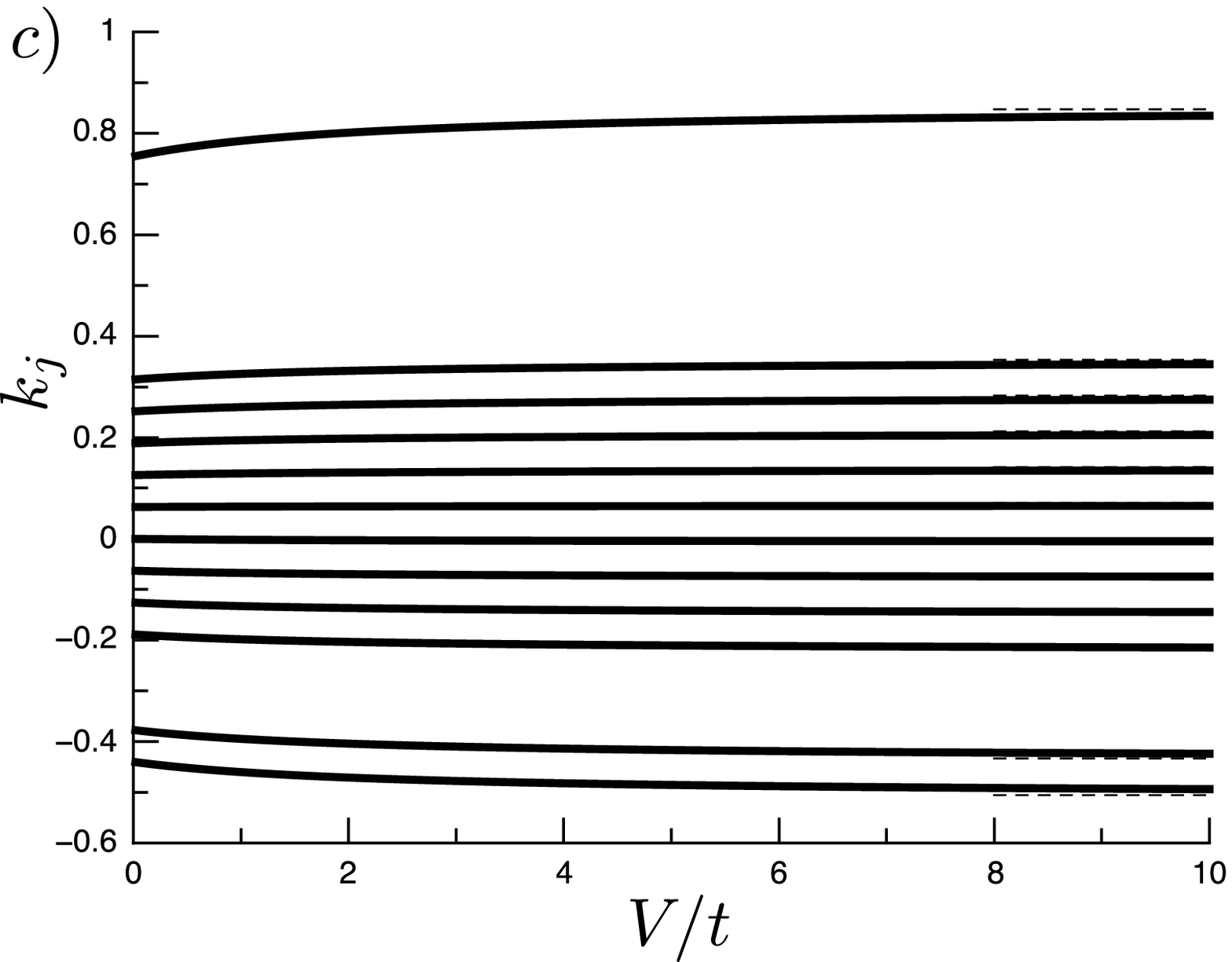}\includegraphics[width=0.5\columnwidth]{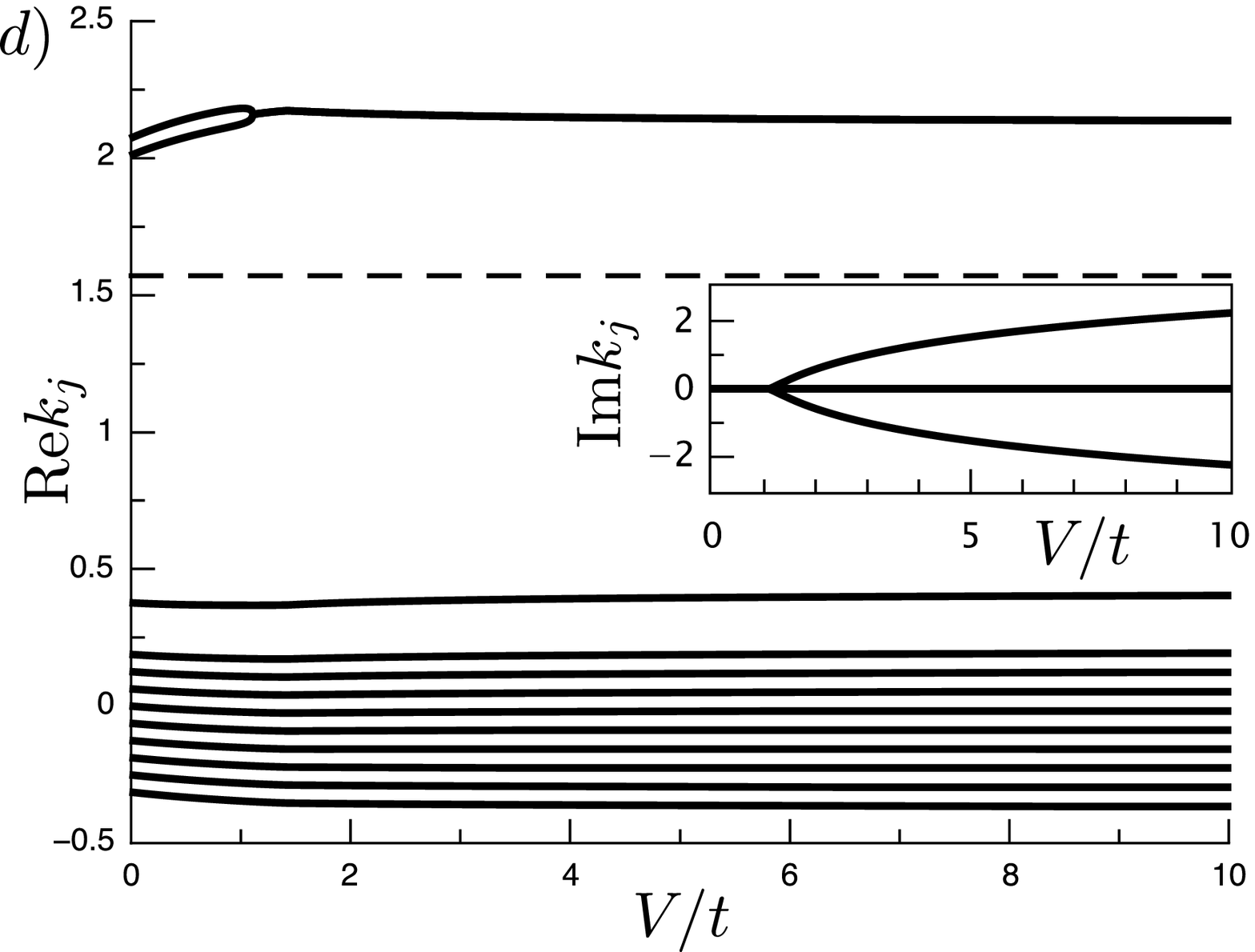}

\caption{The numerical solutions of Bethe equations, Eq. (5) of the main text,
for $k_{j}$ as a function of interaction strengths $V$ for $n=12$
particles and $L=100$ - full lines. The thin dashed lines at large
$V$ correspond to asymptotes from Eq. (7) of the main text. The states
are classified according to Eq. (6) of the main text: a) $\lambda_{j}=\left\{ -5,-4,-3,-2,-1,0,1,2,3,4,10\right\} $;
b) $\lambda_{j}=\left\{ -6,-4,-3,-2,-1,0,1,2,3,4,9\right\} $; c)
$\lambda_{j}=\left\{ -7,-6,-3,-2,-1,0,1,2,3,4,12\right\} $; d) $\lambda_{j}=\left\{ -5,-4,-3,-2,-1,0,1,2,3,6,32,33\right\} $
- a bound state forms out of a pair of quasimomenta with $k_{j}>/\pi/2$
above a finite value of $V$, thick dash line marks the value of $k=\pi/2$,
the inset is the imaginary parts of all quasimomenta $k_{j}$.}
\end{figure}
\begin{figure}[H]
\centering\includegraphics[width=0.5\columnwidth]{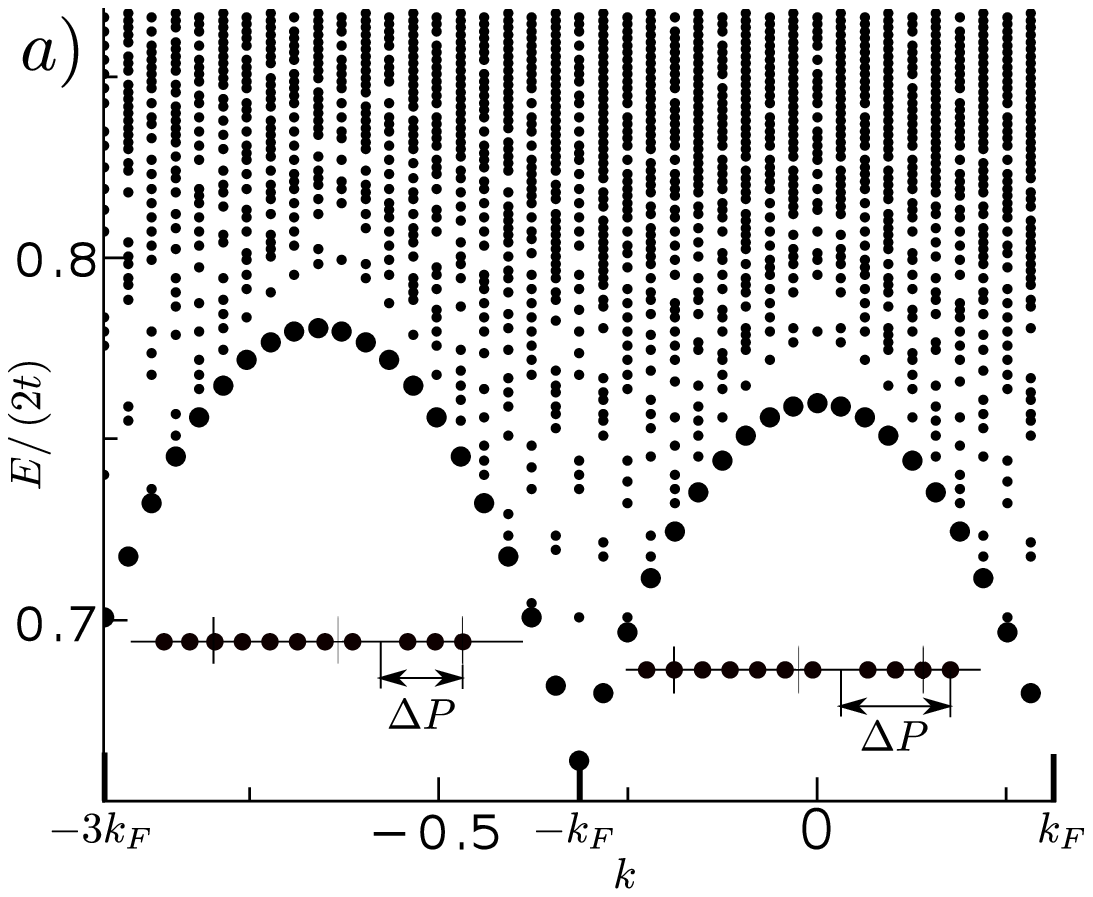}\includegraphics[width=0.5\columnwidth]{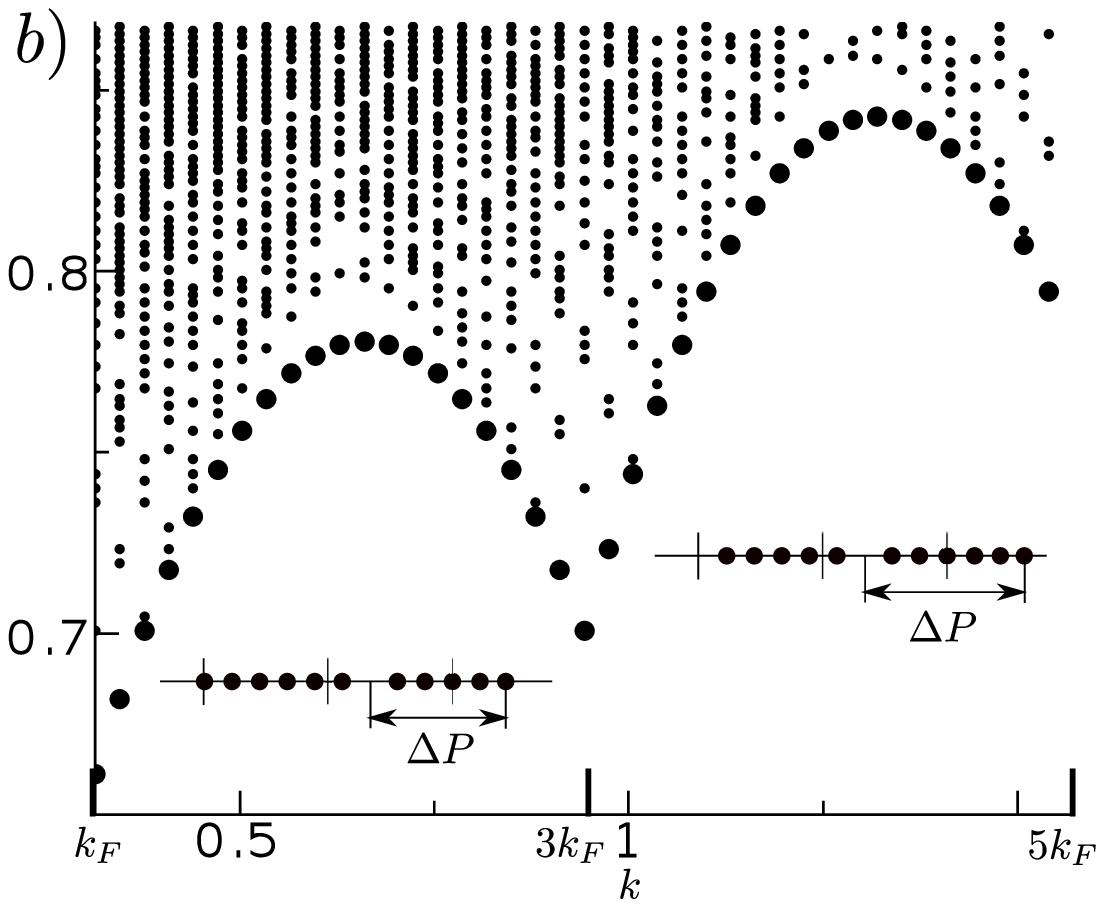}

\includegraphics[width=0.5\columnwidth]{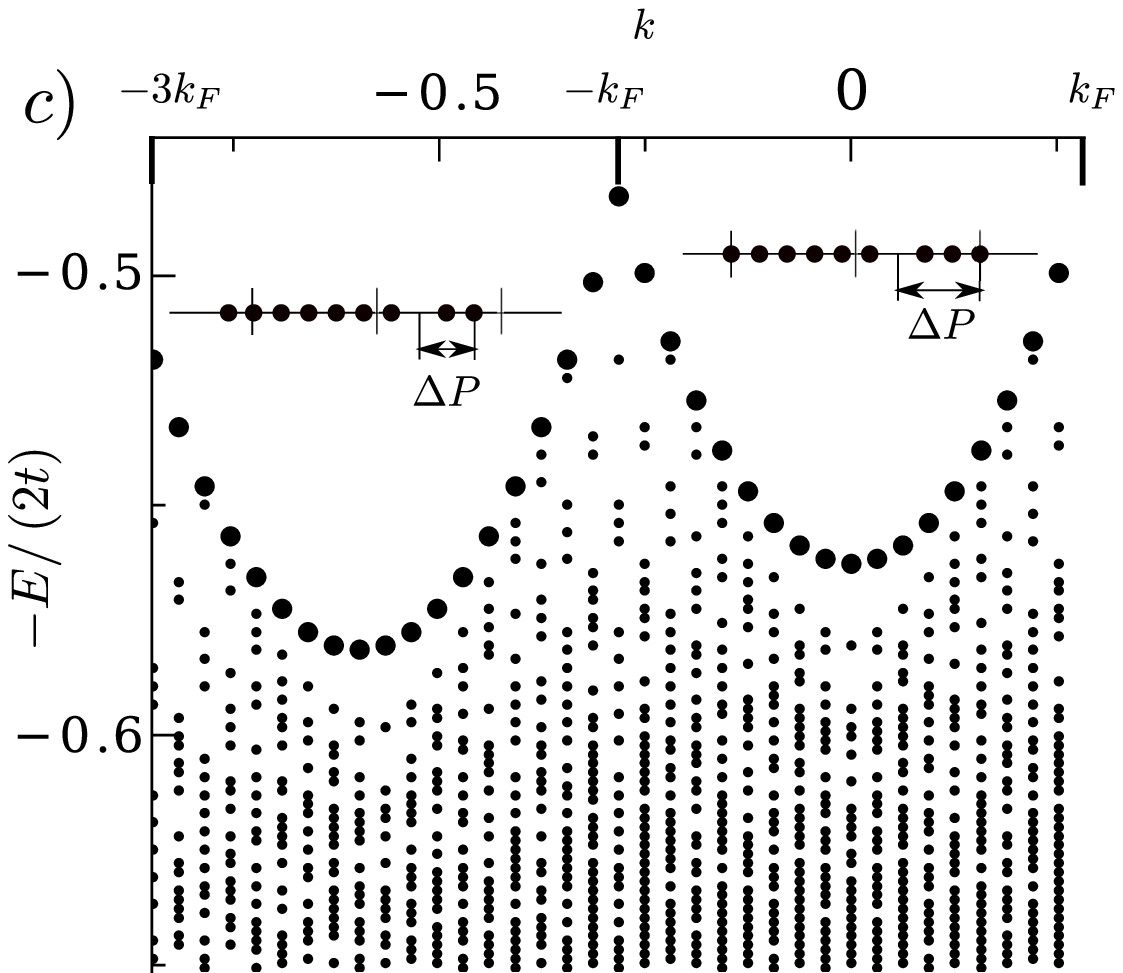}\includegraphics[width=0.5\columnwidth]{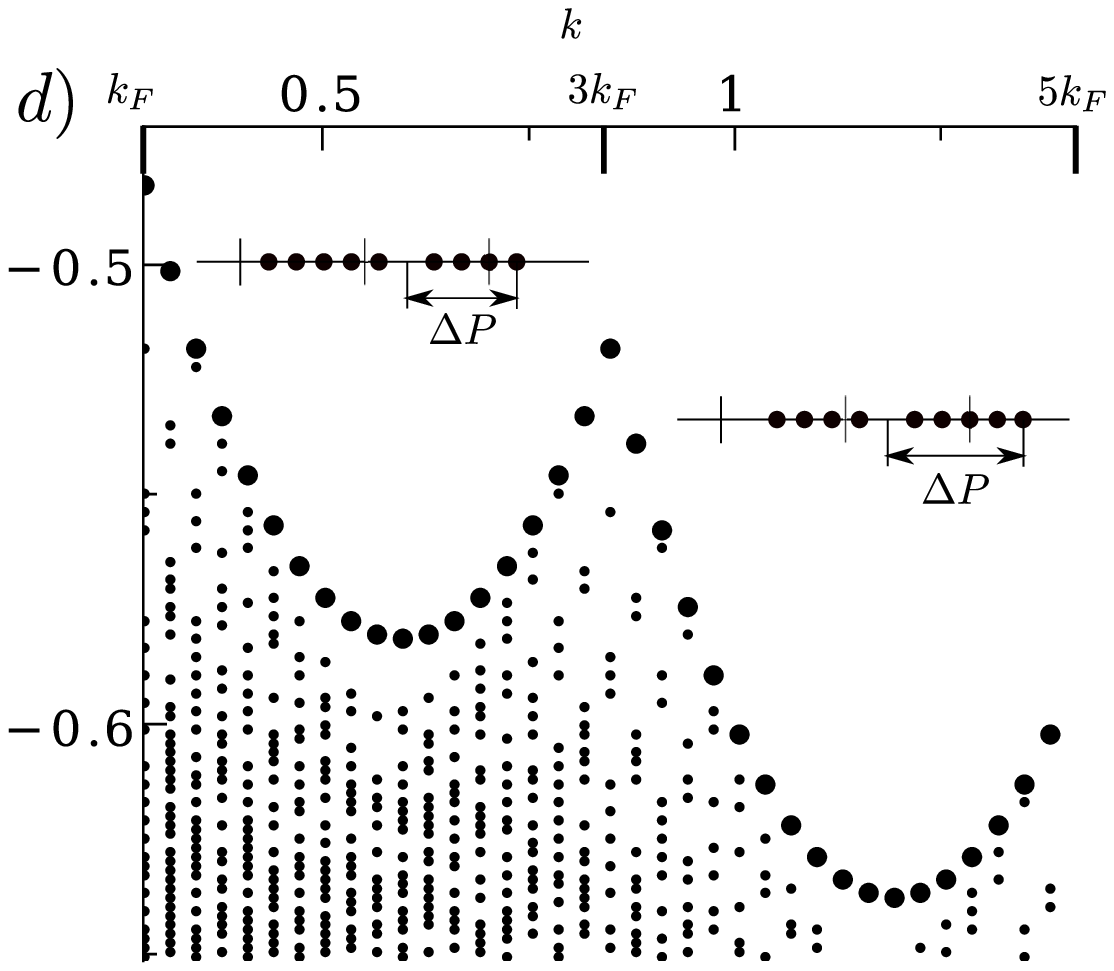}

\caption{Spectrum of the eigenstates, Eq. (3) of the main text, that are involved
in the form factor in Eq. (2) of the main text for the ground state
with $n=19$ particles, $L=200$, and $V\ll t$ - dots. Large dots
are the states at the edge. The insets are sketches of sets of quasimomenta
that correspond to the edge states using classification Eq. (6) of
the main text. Positive half-plane $E-E_{0}>\mu$ are the states with
an extra added particle: a) momenta are from $-3k_{F}$ to $k_{F}$,
the energies at the edge at $-3k_{F}$ and $-k_{F}$ correspond to
chemical potentials $\mu_{-1}$ and $\mu_{0}$; b) momenta are from
$k_{F}$ to $5k_{F}$, the energies at the edge at $k_{F}$, $3k_{F}$,
and $5k_{F}$ correspond to chemical potentials $\mu_{1}$, $\mu_{2}$,
and $\mu_{3}$. Negative half-plane $E_{0}-E<\mu$ are the states
with one particle removed: c) momenta are from $-3k_{F}$ to $k_{F}$;
d) momenta are from $k_{F}$ to $5k_{F}$.}
\end{figure}
\twocolumngrid


\begin{thebibliography}{10}
\bibitem{ContinuumRegime} In the thermodynamic limit the length of
the system $L$ and the number of particles $n$ are infinite but
the density $n/L$ is finite. Small occupancy of the lattice is the
long wave length limit compared to the lattice parameter, $n\ll L$.
\bibitem{hbara}In our notations $\hbar$ and lattice parameter are
equal to one.
\end{thebibliography}
\end{document}